# Broadband Optomechanical Sensing at the Thermodynamic Limit

Feng Zhou, Yiliang Bao, Ramgopal Madugani, David A. Long, Jason J. Gorman[*], and Thomas W. LeBrun[*]

*National Institute of Standards and Technology, Gaithersburg, MD, USA*

[*]e-mail: thomas.lebrun@nist.gov, gorman@nist.gov

**Cavity optomechanics has opened new avenues of research in both fundamental physics and precision measurement by significantly advancing the sensitivity achievable in detecting attonewton forces[1], nanoparticles[2], magnetic fields[3], and gravitational waves[4,5]. A fundamental limit to sensitivity for these measurements is energy exchange with the environment as described by the fluctuation-dissipation theorem. While the limiting sensitivity can be increased by increasing the mass or reducing the damping of the mechanical sensing element, these design tradeoffs lead to larger detectors or limit the range of mechanical frequencies that can be measured, excluding the bandwidth requirements for many real-world applications. We report on a microfabricated optomechanical sensing platform based on a Fabry-Pérot microcavity and show that when operating as an accelerometer it can achieve nearly ideal broadband performance at the thermodynamic limit (Brownian motion of the proof mass) with the highest sensitivity reported to date over a wide frequency range (314 nm·s$^{-2}$/√Hz over 6.8 kHz). This approach is applicable to a range of measurements from pressure and force sensing to seismology and gravimetry, including searches for new physics such as non-Newtonian gravity[6] or dark matter[7-9].**

High-precision, high-bandwidth acceleration measurement is central to many important applications, including inertial navigation[10,11], seismometry[12,13], and structural health monitoring of buildings and bridges[14]. Traditional electromechanical accelerometers have largely relied on piezoelectric, capacitive, or piezoresistive transduction to convert the displacement of the accelerometer's proof mass to an output voltage when an excitation is applied. However, these transduction methods have reached sensitivity and bandwidth limits that are prohibitive for many applications. As a result, optical accelerometers have long been of interest due to the high precision provided by interferometry. These have included accelerometers assembled from macroscale optics[15], as well as those based on fiber optic interferometers[16] and fiber Bragg grating cavities[17]. More recently, the development of integrated micro- and nanoscale optomechanical devices has produced accelerometers with significantly better acceleration sensitivity than previously reported. Examples include a zipper photonic crystal optomechanical cavity in silicon nitride[18] and a fiber-based microcavity integrated into a fused silica mechanical resonator[19,20]. These integrated micro- and nanoscale cavities provide displacement sensitivity in the range of 1 fm/√Hz and below due to their low optical loss, which can result in an acceleration sensitivity on the order of 1 μm·s$^{-2}$/√Hz and below for frequencies of oscillation up to 10 kHz or more[18-23].

In addition to high sensitivity, optomechanical accelerometers promise greater accuracy without the need for calibration because the displacement of the proof mass can be measured directly in terms of the laser wavelength—the definition of length in the International System of Units[24]—rather than electrical quantities. However, the device physics must be accurately known to determine the acceleration from the displacement measurement. Therefore, the accelerometer must have a simple, deterministic mechanical response so that the dynamic model can be accurately inverted to convert displacement to acceleration. Ideally, the thermomechanical noise of the accelerometer should exceed the other fundamental noise source, optical shot noise in the displacement measurement, so that the mechanical response can be identified with high fidelity and the acceleration noise floor will be flat over a wide frequency range[25-27].

In previous work, the mechanical mode structure has been too complex and difficult to identify to allow reliable, broadband conversion between displacement and acceleration, or shot noise has dominated over most of the bandwidth of the accelerometer, or both, thereby preventing broadband measurement. In this letter, we demonstrate a microfabricated optomechanical accelerometer that reaches the thermodynamic limit of sensitivity broadband (314 nm·s$^{-2}$/ $\sqrt{}$ Hz over 6.8 kHz), greatly exceeding the resolution and bandwidth found in conventional accelerometers. Broadband measurement is necessary for detection of general time varying signals at the thermodynamic limit, as well as rigorous understanding of the device physics required for advanced applications. In addition, the devices reported here are fully packaged, field-deployable, operable in air and vacuum—and achieve the highest acceleration sensitivity reported to date for a compact optomechanical accelerometer. While the specific measurement discussed here is acceleration, the platform is applicable to sensing as well as fundamental physical measurements of small forces such as departures from Newtonian gravity at small distances[6] or array-based searches for dark matter through gravitational force detection[7-9].

The design of the optomechanical accelerometer and its components are described in Fig. 1. Two silicon microfabricated chips comprise the main sensing elements of the accelerometer. One chip has a mechanical resonator and the other has a concave micromirror, both with patterned dielectric mirror and anti-reflective coatings. A hemispherical Fabry-Pérot cavity is formed by assembling the chips such that the displacement of the mechanical resonator can be measured with high precision by interrogating one of the cavity's optical resonances. When an acceleration is applied to the accelerometer package, that signal is transduced by measuring the displacement of the mechanical resonator and converting it to a measured acceleration. The mechanical resonator is composed of a single-crystal silicon proof mass (thickness: 525 μm, width: either 3 mm or 4 mm square, mass: approximately 11 mg or 20 mg) that is constrained on both sides by 1.5 μm thick silicon nitride beams to ensure nearly ideal piston-like displacement in response to an acceleration perpendicular to the chip's surface. The concave micromirror is fabricated in single crystal silicon using a wet etching process[28,29], resulting in high-quality mirrors with radii of curvature of approximately 410 μm, a depth of 257 μm, and a surface roughness of 1 nm RMS. The two chips are assembled together and bonded with UV curable adhesive, and the resulting chip stack is assembled in a stainless-steel package that includes a polarization-maintaining fiber and a custom fiber focusing lens for mode coupling to the cavity (see Methods for more details). This sensor design can be extended to a range of measurements such as force, pressure, seismology and gravimetry by simply modifying the mechanical resonator to have the appropriate mass, stiffness, and damping properties for the given application.

The optical spectrum of the hemispherical cavity was measured in both transmission and reflection, as shown for wavelengths near 1550 nm in Fig. 2, where the free spectral range (FSR) is 400 GHz (3.21 nm), the coupling efficiency is 80 %, and higher-order transverse modes can be seen between the dominant fundamental modes. These modes were imaged in transmission on an InGaAs camera, showing intensity profiles characteristic of highly symmetric spatial modes. Displacement measurements of the mechanical resonator were performed in reflection using a fundamental cavity mode ($TEM_{00}$) near a wavelength of 1551 nm with a linewidth of $\Gamma$ = 73.7 MHz (FWHM), a finesse of $F$ = 5430, and a mirror reflectivity of $R$ = 99.89 %, as shown in Fig. 2b. The selection of $F$ was based on the tradeoff between sensitivity and dynamic range for a sidelock-based measurement.

The readout method used for small-amplitude displacement measurement of the optical cavity is shown in Fig. 3a, where a highly stable fiber laser (FL) is locked at a fixed reflected intensity on the side of the optical resonance by tuning the wavelength of the laser using an electro-optic modulator (EOM). Side-locking is achieved with a low bandwidth controller ($\approx$ 300 Hz) so that slow changes in cavity length are tracked by the laser wavelength while faster motion of the mechanical resonator generates intensity fluctuations that are used to detect acceleration (see Methods). This approach is applied to measure the displacement spectral densities in Fig. 3 due to the superior broadband noise performance of the FL. In addition, a widely tunable external cavity diode laser (ECDL) was used for the data in Figs. 2 and 4 due to its wider wavelength range and resulting ability to easily tune to a desired cavity mode under rapidly varying measurement conditions (see Supplementary Information).

The displacement sensitivity was measured in air and in vacuum at room temperature while the accelerometer was acoustically and vibrationally isolated. The resulting displacement spectral density in air is shown in Fig. 3b, where a single vibrational mode is present between 100 Hz and 28 kHz. This is the first demonstration reported of an optomechanical accelerometer operating with a single vibrational mode over such a wide bandwidth. Pure single-mode response is essential to enable accurate conversion of displacement to measured acceleration from first principles.

A fit of the displacement spectral density to the thermomechanical response for a simple harmonic oscillator shows close agreement in Fig. 3b (see Supplementary Information), allowing precise estimates of the resonance frequency, $\omega_0 = 2\pi*9.852(16)$ kHz, quality factor, $Q$ = 99.4(1.8), and mass, $m$ = 10.8(9) mg. This mass estimate derived from the thermomechanical fit is well within the uncertainty of the value of 11.01(53) mg calculated from the dimensions of the silicon resonator and optical coatings (see Methods). The noise floor at the lowest frequencies is set by readout noise that is likely due to laser frequency noise, phase modulation noise from the EOM, or thermal effects. Well above resonance, approaching 28 kHz, the noise floor closely approaches the optical shot noise limit.

Importantly, the displacement sensitivity is limited by thermomechanical noise over most of the measured frequency range. This was achieved by optimizing the optical ($L$, $F$) and mechanical ($m$, $Q$, $\omega_0$) parameters so that the thermomechanical noise is above or equal to the shot noise within the bandwidth of interest. An additional benefit of being broadband limited by thermomechanical noise is that the harmonic oscillator model fit can be very accurate due to a high signal-to-noise ratio, which provides greater precision when converting from proof mass displacement to acceleration.

Comparing the displacement spectral density in air and vacuum in Fig. 3c, the increased $Q$ in vacuum, due to a reduction in gas damping, results in larger thermomechanical noise on resonance and less away from resonance, as expected. However, due to the balance between the thermomechanical noise and shot noise, the frequency range over which the spectral density is thermomechanically limited is clearly reduced. The displacement spectral densities in Fig. 3c are converted to a noise equivalent acceleration (NEA) by dividing the response by the harmonic oscillator transfer function (see Supplementary Information), as shown in Fig. 3d. As expected, the NEA reaches the acceleration thermomechanical limit set by the Langevin force ($a_{th} = \sqrt{4k_B T \omega_0 / mQ}$, see Supplementary Information) wherever the displacement spectral density is limited by thermomechanical noise. The Langevin force is reduced when the damping is lower, providing a lower thermodynamic limit but making it more difficult to reach since the shot noise must be lower than the thermomechanical noise. Due to increased damping in air, the minimum NEA is higher, 912 nm·s$^{-2}$/√Hz (93 ng$_n$/√Hz, 1 g$_n$ = 9.81 m·s$^{-2}$), than in vacuum, 314 nm·s$^{-2}$/ √Hz (32 ng$_n$/√Hz), where the vacuum sensitivity represents the lowest value reported for a compact, low-mass optomechanical accelerometer. The bandwidth over which the NEA is within 3 dB of the acceleration thermomechanical limit is 13.6 kHz and 6.8 kHz for air and vacuum, respectively. This wide range is made possible by the exceptionally low displacement readout noise. Furthermore, the NEA only varies by one order of magnitude over the frequency range, which is an improvement of two to four orders of magnitude compared to previously reported optomechanical accelerometers. This reasonably flat NEA is essential for making high-precision broadband acceleration measurements since it provides a consistent signal-to-noise ratio over the measurement bandwidth.

As a test of sensing performance for a range of external accelerations, the optomechanical accelerometer was placed on a piezoelectric shaker table and the accelerometer output was compared with the motion measured with a homodyne Michelson interferometer (see Fig. 4a and Methods). The frequency of the sinusoidal acceleration generated by the shaker was swept from 1 kHz to 20 kHz. The interferometer was used to measure the displacement of the accelerometer package, which has a 5 mm square gold-on-silicon mirror bonded to it. The resulting displacement amplitude as a function of drive frequency is shown in Fig. 4b, where the displacement of the proof mass and package are different because the accelerometer response includes the resonance of the proof mass (9.86 kHz) and the first resonance of the shaker (12.68 kHz), whereas the external interferometer can only detect the shaker resonance. The inset shows shaker linearity that is better than 1.3 % (see Supplementary Information). In addition to the large resonances, much smaller structures in the accelerometer displacement data can be seen at 3.9 kHz and 11.6 kHz. They have been linked to the accelerometer packaging and the shaker itself and are dependent on the torque used in mounting the accelerometer onto the shaker.

The displacement data from the accelerometer was converted to acceleration and the interferometer displacement data was transformed to acceleration by multiplying by $(2\pi f_d)^2$, where $f_d$ is the drive frequency. As shown in Fig. 4c, there is close agreement between the accelerometer and interferometer throughout the entire 20 kHz bandwidth. The accelerometer's fundamental resonance disappears in the acceleration data due to the model inversion, demonstrating that measurement on and even above resonance can be effective for these single-mode devices at the thermal limit. The percent deviation of the accelerometer from the interferometer was calculated at each measurement frequency. The standard deviation of this value over the entire frequency range is 15.9 % and between 4.5 kHz and 11 kHz it is 9.7 % after applying a moving average filter to the interferometer data to reduce noise (see Supplementary Information). This comparison helps confirm that the accelerometer is behaving like a harmonic oscillator (i.e., exhibiting a single, one-dimensional, viscously-damped, piston mode of the proof mass), but does not accurately indicate accelerometer performance, as the deviation is dominated by the mechanics of the external reference interferometer. This represents the widest bandwidth demonstrated to date at this error level using a first-principles description based on a single degree-of-freedom oscillator.

Selecting a mechanical resonator with low $Q$ and reasonably high mass was necessary to set the thermomechanical noise to be equal to or larger than the shot noise while still achieving a minimum NEA of 32 ng/√Hz. This approach is contrary to the widely accepted approach of maximizing $Q$ in order to minimize the acceleration thermomechanical noise. The apparent contradiction is due to the combined effects of the thermomechanical noise and the optical shot noise (see Supplementary Information). Increasing $Q$ is only beneficial for broadband measurements if the shot noise can be reduced below the thermomechanical noise. In addition to the frequency independent NEA and the accurate fit to the thermomechanical model shown in Fig. 3, low $Q$ has two other important advantages for optomechanical accelerometers: 1) the mechanical ring-down time is proportional to $Q$, and 2) it has been shown that the linear dynamic range of a flexural mechanical resonator is inversely proportional to $Q$[30,31]. Therefore, a low-$Q$ accelerometer is able to respond more quickly to transient broadband excitations while capturing a wider range of acceleration amplitudes without demonstrating mechanical nonlinearity. Noting the importance of low $Q$ for broadband measurement, a better approach for reducing the thermomechanical noise, assuming that the shot noise can be reduced as well, is to increase the proof mass (and increase the stiffness if a fixed resonance frequency is needed.)

In conclusion, we have demonstrated a compact optomechanical accelerometer that achieves the thermodynamic limit of sensitivity over a frequency range greater than 15 kHz, including on, above and below resonance. In contrast to previous work, the limiting sensitivity can be achieved for general signals and the highly ideal single-mode structure enables accurate inversion of the mechanical response for accurate measurement. Additionally, broadband measurement at the thermodynamic limit yields a detection sensitivity nearly independent of frequency, so that resonant enhancement is not necessary for detection of weak signals and detection even above resonance is possible with the same noise-equivalent sensitivity despite a rapidly falling response. The compact size of the sensor enables high-precision measurements outside of laboratory settings, and the optomechanical sensing platform is widely applicable to measurements beyond acceleration, such as force, pressure, and gravity sensing, through straightforward modification of the mechanical resonator.

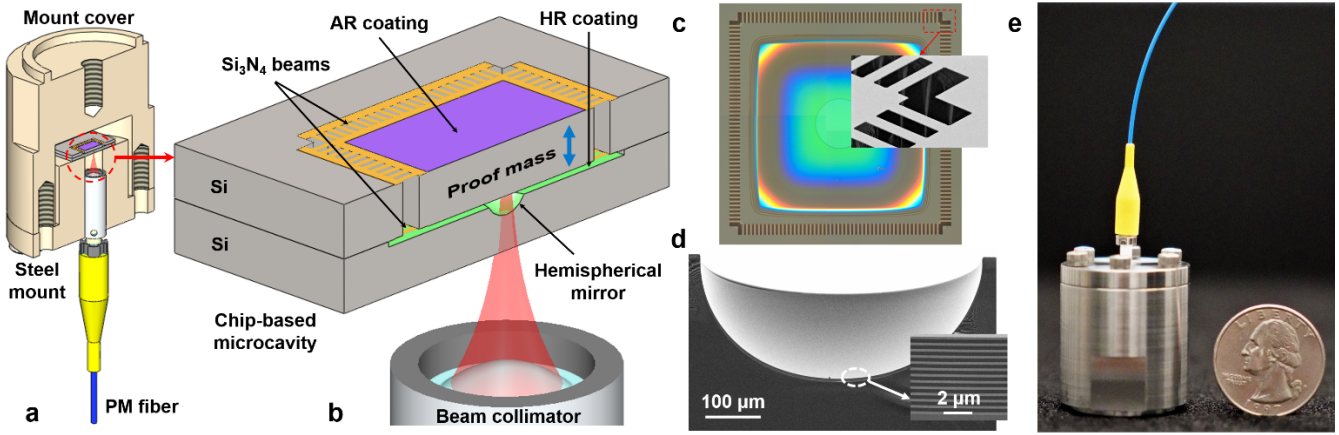

**Fig. 1. Optomechanical accelerometer design. a,** Cross-section of the accelerometer, including microfabricated cavity optomechanical components, fiber optics for coupling light into and out of the optical cavity, and a stainless-steel package. **b,** Cross-section of the two microfabricated chips, in which one chip contains a mechanical resonator composed of a millimeter-scale silicon proof mass suspended by silicon nitride ($Si_3N_4$) microbeams on both sides and the other chip has a concave silicon micromirror. The proof mass and micromirror form a hemispherical optical cavity, where both chips have high-reflectivity and anti-reflective dielectric coatings. Motion of the proof mass resulting from an acceleration of the package can be measured in real time by tracking the cavity length. **c,** Stitched optical micrograph of the mechanical resonator showing the high-reflectivity mirror coating restricted to the proof mass to avoid fouling the microbeams. Inset: Scanning electron micrograph of the silicon nitride microbeams. **d,** Scanning electron micrograph of a cleaved concave silicon micromirror. Inset: Close-up of the high-reflectivity mirror coating with quarter-wave periodicity. **e,** Image of a packaged and fiber-coupled accelerometer.

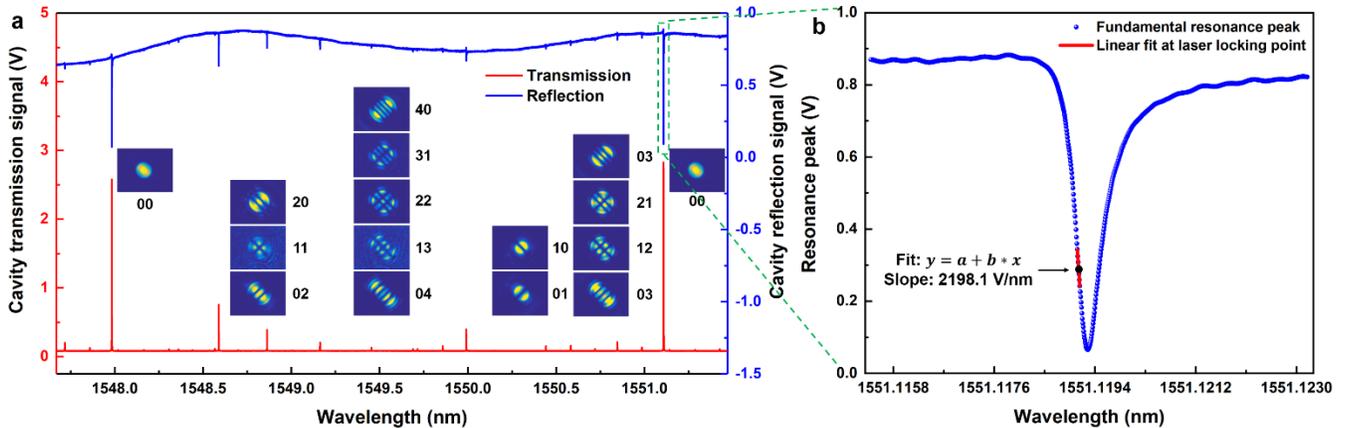

**Fig. 2. Spectra for the optical cavity. a,** Reflected and transmitted spectra for the optical cavity over a single free spectral range (FSR) near 1550 nm. Higher-order transverse modes in addition to the fundamental ($TEM_{00}$) modes are imaged in transmission using an InGaAs camera. **b,** A single fundamental mode that is used to transduce the motion of the proof mass is shown, where the optical finesse $F$ is 5430. The red region on the resonance indicates the location for side-locking to the cavity.

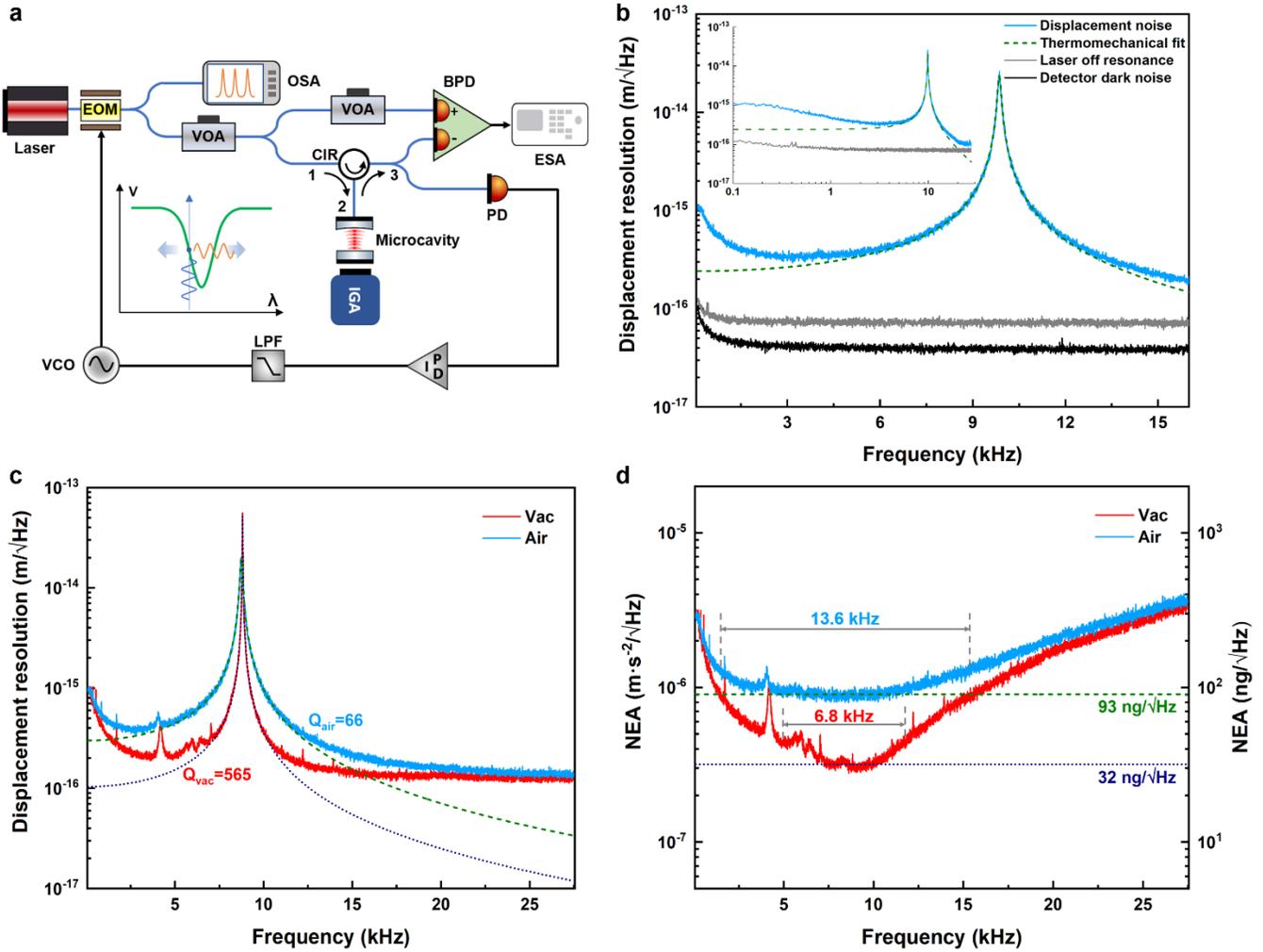

**Fig. 3. Displacement spectral densities and the noise equivalent acceleration. a,** Diagram of the optical cavity readout method used to measure the noise performance of the accelerometer. A highly stable fiber laser is phase modulated with an electro-optic modulator (EOM) for wavelength tunability. A single sideband on the laser is tuned to the optical resonance and locked to the point of highest slope (i.e., highest sensitivity) using a low-bandwidth ($\approx$ 300 Hz) servo controller. The intensity fluctuations resulting from motion of the proof mass are measured using balanced photodetection (BPD) to mitigate intensity noise, and an electronic spectrum analyzer (ESA). VOA: variable optical attenuator, OSA: optical spectrum analyzer, VCO: voltage-controlled oscillator, CIR: circulator, IGA: InGaAs camera, PD: photodetector, LPF: low-pass filter, PID: proportional-integral-derivative controller. **b,** Displacement spectral density for the accelerometer in air, showing a single vibrational mode. Dashed line: Fit to the thermomechanical noise model. Grey line: Shot noise when the laser sideband is completely detuned from the optical resonance. Black line: Photodetector dark noise. Inset: Log-log plot of displacement spectral density, showing the purity of the single vibrational mode over a wide frequency range. **c,** Comparison between operation in air and in vacuum, $P$ = 133 mPa (1 mTorr). Dashed lines: Respective fits to the thermomechanical noise model. **d,** Noise equivalent acceleration (NEA) found by converting the displacement in Fig. 3c to acceleration using the harmonic oscillator model (see Supplementary Information). Indicated frequency bands represent the range over which the NEA is within 3 dB of the acceleration thermomechanical noise limit (dashed lines).

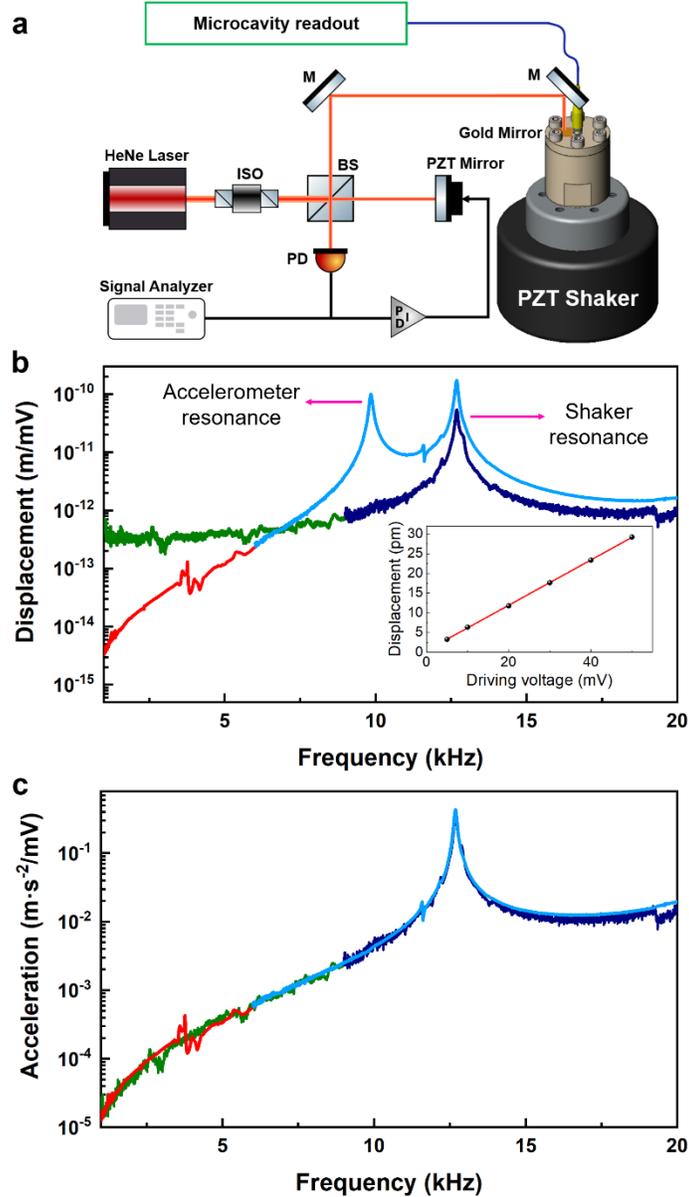

**Fig. 4. Shaker table testing of the accelerometer. a,** Experimental configuration for the shaker table tests, showing the combination of a piezoelectric shaker table, the optomechanical accelerometer, and a homodyne Michelson interferometer. The interferometer probes the back of the accelerometer package, providing a direct measurement of the excitation experienced by the accelerometer package. Mirror (M), photodetector (PD), non-polarizing beamsplitter (BS), optical isolator (ISO), proportional-integral-derivative servo loop (PID). The microcavity readout is shown in Fig. 3a. **b,** Comparison of the displacement measured with the accelerometer and interferometer. **c,** Acceleration measured by the accelerometer and interferometer after converting the displacement data in (b), showing close agreement. The displacement sensitivity of the accelerometer is approximately 600 times greater than the interferometer (0.1 fm/√Hz and 60 fm/√Hz, respectively). As a result, different drive voltages were used in the shaker table tests to get above the noise floor of the interferometer while not perturbing the laser lock for the accelerometer, 0.1 mV (blue) and 25 mV (red) for the accelerometer and 5 mV (navy) and 30 mV (green) for the interferometer, respectively. The shaker was found to be highly linear for this drive voltage range (see inset in (b) and Supplementary Information), making this comparison possible.

**Data Availability**

**Acknowledgements**
The authors thank Ben Reschovsky for helpful discussions and sharing analysis of the uncertainties derived from fits to thermomechanical spectra. This work was partially supported by the NIST on a Chip Program. Y.B. acknowledges support from the National Institute of Standards and Technology (NIST), Department of Commerce, USA (70NANB17H247). This research was performed in part in the NIST Center for Nanoscale Science and Technology NanoFab.

**Contributions**
J.G. and T.L. conceived the research. J.G., T.L and Y.B. designed the sensor. Y.B. microfabricated the sensors. F.Z. and Y.B. assembled the sensors. F.Z. measured optical spectra. F.Z., D.L. and T.L. designed the optical readout. F.Z., D.L. and R.M. measured displacement spectral densities. F.Z., Y.B. and J.G. performed shaker table testing. All authors analyzed the data and wrote the paper.

**Corresponding authors**
Correspondence may be addressed to T. LeBrun or J. Gorman.

**Competing interests**
The authors declare no competing interests.

**Methods**

**Device Geometry and Mounting.** Measurements on two optomechanical accelerometers are reported here. The two differ principally in the size of the proof mass and the packaging. Device A is operated in air for noise measurements, Fig. 3b, and on the shaker for driven frequency response measurements, Fig. 4. It is shown in Fig. 1 and has a 3 mm x 3 mm x 0.525 mm proof mass and a resonant frequency of 9.86 kHz. Device B is a bare device mounted without a cover for vacuum compatibility (Figs. 3c and 3d) and has a 4 mm x 4 mm x 0.525 mm proof mass and a resonant frequency of 8.74 kHz. Device A is operated in air on a mechanical isolation stage inside an acoustic shield. Device B is operated in a vacuum chamber at varying pressure on a pneumatic optical table.

**Resonator Mass.** The value of the proof mass in the mechanical resonator was calculated using the designed geometry and approximate densities for single-crystal silicon and the optical coatings, resulting in 11.07(53) mg for Device A and 19.59(94) mg for Device B. The main source of uncertainty in the mass is the variation in the silicon wafer thickness (±25 μm) which gives a relative uncertainty of approximately 5% for the calculated mass. This only limits *a priori* estimate of the mass, not the uncertainty of the acceleration measurement, which relies on *in situ* measurement of $\omega_0$ and $Q$.

A similar proof mass from the same fabrication process was measured for Devices A and B after being removed from the chip. The masses were calibrated by the NIST Mass and Force Group and found to be 11.13 mg for Device A and 19.88 mg for Device B, which deviate from the calculated value by 0.5 % and 1.5 %, respectively. Any microbeams adhering to the proof mass after removal would increase the mass by less than 20 μg, and the uncertainty of the calibrated values[32] is also negligible relative to the uncertainty of the calculated values.

**Uncertainties in Parameters Estimated from Fits.** Fitting thermomechanical noise spectra allows $\omega_0$, $Q$, and $m$ to be measured, given the temperature. These values can vary over time due to changes in laboratory conditions, such as temperature, aging from sources including curing of packaging adhesive or accumulated stress from cycling between air and vacuum. To estimate the associated uncertainties, we use the standard deviation of multiple measurements on a device over a period of approximately eleven months. The uncertainty reported by the fitting routines is not included in the stated uncertainty as it is small compared to the variation over a year, even when accounting for variation in fitting procedures. This represents a conservative estimate for the measurements reported here. The uncertainty can be substantially reduced, for example by measuring $\omega_0$ and $Q$ immediately before and after acceleration measurement, but best practice for accurate acceleration metrology with the devices is outside the scope of this work and will be reported elsewhere. For Device A the relative uncertainties for $\omega_0$, $Q$, and $m$ are approximately 0.2%, 2%, and 8%, respectively. Only the uncertainties in $\omega_0$ and $Q$ directly contribute to the uncertainty in acceleration measurement.

**Accelerometer Fabrication and Assembly.** The accelerometer is composed of two silicon chips bonded together, one with a concave silicon micromirror and the other with a mechanical resonator including the proof mass. The concave silicon micromirror was fabricated using a slow isotropic wet etching process on a double-side polished, 525 μm thick silicon wafer. First, a 35 μm deep recess was etched using deep reactive ion etching (DRIE), providing space between the moving proof mass and micromirror. Then, the wafer was coated with stochiometric silicon nitride (300 nm thick) using low pressure chemical vapor deposition (LPCVD), which serves as a hard mask during wet etching. Circular apertures 300 μm in diameter were patterned in the silicon nitride layer using reactive ion etching (RIE). The wafer was then etched in a mixture of hydrofluoric, nitric, and acetic acids (HNA, 9:75:30 ratio) at room temperature for a predetermined time to achieve the desired depth and radius of curvature, which are approximately 257 μm and 410 μm, respectively, in the presented accelerometers. Additional fabrication details have been published previously[29]. A protective coating was applied to the wafer before dicing and the resulting 1 cm chips were then cleaned.

The mechanical resonator was fabricated on a double-side polished, 525 μm thick silicon wafer by patterning both sides of the wafer identically. A 1.5 μm thick, low-stress silicon nitride layer was deposited on the wafer using LPCVD. The proof mass and beam geometry were patterned with optical lithographically and the silicon nitride was etched with RIE. DRIE was then used to etch the beam pattern through the silicon wafer from both sides in subsequent etch steps. A protective coating was applied to the wafer before dicing and the resulting 1 cm chips were then cleaned. Finally, the beams and proof mass were released at the chip level by undercutting the silicon nitride beams using KOH with a concentration of 30 % at 60 °C.

Dielectric mirror and anti-reflection coatings with alternating tantalum pentoxide and silicon dioxide layers were applied to the concave micromirrors and mechanical resonators using ion beam sputtering. A shadow mask made from an etched silicon wafer was used to selectively deposit the coatings on the proof mass and concave mirror. A pair of the completed chips were aligned and bonded with UV curable adhesive. This is a self-aligned process that requires no adjustment of angle or translation beyond ensuring overlap of the concave micromirror and proof mass. Finally, the chip assembly was aligned to the fiber collimator and bonded to the accelerometer package using UV curable adhesive. Anti-reflection coatings on the focusing lens and the back of the proof mass are used to reduce parasitic reflections

**Optical Interrogation.** In order to interrogate the accelerometer, a rack mounted unit containing three fiber lasers was employed. Each laser is tunable over a range of 1 nm and exhibit a short-term linewidth near 100 Hz (details of the experimental setup are shown in Figure 3). To facilitate side locking of a given fiber laser to the optical cavity, the laser was modulated using an electro-optic phase modulator driven by an amplified voltage-controlled oscillator to generate sidebands near 3 GHz for frequency stabilization. We tuned one sideband to the maximum slope point of the resonance and sent the error signal induced by variations in the reflected optical power to a proportional-integral-derivative controller. This side lock compensated for slow changes in intensity, largely due to thermal or humidity induced drift of the cavity length, while having no effect on intensity fluctuations that exceed the servo control bandwidth (<300 Hz). To suppress laser intensity noise, a balanced detection scheme with a bandwidth near 1 MHz was used. The resulting signal from the balanced detector was digitized using a 12-bit radio-frequency spectrum analyzer with a bandwidth of 28 kHz.

The reflected intensity fluctuations for the side-locked cavity result in a detector voltage, $\Delta V$, that is converted to displacement, $\Delta L$, using the relation $\Delta L = L\, \Delta V/(\lambda\, S)$, where $L$ is the nominal cavity length, $\lambda$ is the nominal cavity resonance wavelength and $S = dV/d\lambda$ is the slope of the optical resonance at the lock point (see details in Supplementary Information).

**Interferometric Measurements with the Shaker Table.** The homodyne Michelson interferometer used to test the accelerometer on a shaker table is shown in Fig. 4a. A 632.8 nm stabilized HeNe laser is split into the measurement and reference arms of the interferometer using a non-polarizing 50/50 beam splitter. The light in the reference arm is reflected off of a piezoelectric-actuated mirror and light in the measurement arm is reflected off of a 5 mm square gold mirror mounted on the optomechanical accelerometer package. The reflected light from both arms interferes on a photodetector. The interferometer is locked to the quadrature point (i.e., point of highest fringe slope) using the piezoelectric mirror in the reference arm and a servo controller with a bandwidth below 100 Hz. Shaker vibrations above the servo bandwidth are measured with the interferometer and are converted to displacement using the measured fringe amplitude and laser wavelength, resulting in a noise floor of approximately 60 fm/√Hz above 1 kHz. The optomechanics for the interferometer sit on the same optical table as the shaker table, making them susceptible to vibrations driven by the shaker, as seen in the data in Fig. 4.

# *Supplementary Information*

# Broadband Optomechanical Sensing at the Thermodynamic Limit

Feng Zhou, Yiliang Bao, Ramgopal Madugani, David A. Long, Jason J. Gorman*, and Thomas W. LeBrun*

*National Institute of Standards and Technology, Gaithersburg, MD, USA*
*e-mail: thomas.lebrun@nist.gov, gorman@nist.gov

## S1. Harmonic Oscillator Model

A major benefit of the accelerometer described in the letter is that its dynamic response closely follows that of a one-dimensional viscously-damped harmonic oscillator, making it possible to convert from measured proof mass displacement to an equivalent acceleration using a low-order model. In this section, we describe the harmonic oscillator model and the conversion between displacement and acceleration. Much of the analysis in this section and the next follows directly from the work of Gabrielson[S1] but is specifically focused towards the optomechanical accelerometer.

The harmonic oscillator model is described in Fig. S1, where a mass-spring-damper system is driven by a base excitation, $x_e$, and a Langevin force, $F_L$, that results in thermomechanical noise. The equation of motion for this model is

$$m\ddot{x} + c(\dot{x} - \dot{x}_e) + k(x - x_e) = F_L \quad \text{(S1)}$$

where $m$ is the mass, $k$ is the spring stiffness, $c$ is the damping coefficient, and $x$ is the displacement of the mass. Defining the change in optical cavity length, $x_c$, as $x_c = x - x_e$ and the base acceleration, $a_e$, as $a_e = \ddot{x}_e$ results in the model of interest:

$$\ddot{x}_c + \frac{\omega_0}{Q}\dot{x}_c + \omega_0^2 x_c = -a_e + \frac{F_L}{m} \quad \text{(S2)}$$

where $\omega_0 = \sqrt{k/m}$, $\omega_0 = 2\pi f_0$, $f_0$ is the resonance frequency in the absence of damping, $Q = m\omega_0/c$, and $Q$ is the quality factor.

The relationship between cavity displacement, $x_c$, and base acceleration, $a_e$, as a function of frequency, $\omega$, can be determined from eq. (S2) by neglecting the Langevin force, $F_L$.

$$x_c(\omega) = \frac{-1}{\omega_0^2 - \omega^2 - i\frac{\omega_0 \omega}{Q}} a_e(\omega) = G(i\omega) a_e(\omega) \quad \text{(S3)}$$

The amplitude of $a_e$ can then be written as

$$|a_e(\omega)| = |G(i\omega)|^{-1} |x_c(\omega)|, \quad \text{(S4)}$$

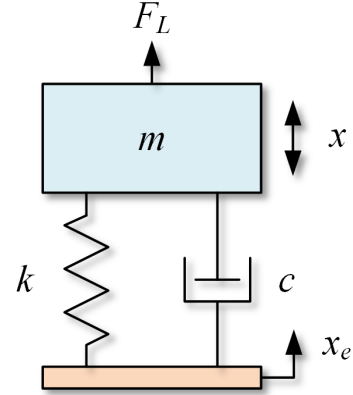

**Fig. S1 Harmonic oscillator model described by a mass-spring-damper system.** $m$: mass, $k$: spring stiffness, $c$: damping coefficient, $x$: proof mass displacement, $x_e$: base displacement, $F_L$: Langevin force.

which has been used to generate the data in Figs. 3d and 4d in the letter. Implementing eq. (S4) requires measurement of $\omega_0$ and $Q$. Here, this was done by applying a least-squares fit of $|G(i\omega)|$ to the data in Figs. 3b and 3c in the letter. A more general conversion between $x_c$ and $a_e$ can be implemented in the time domain using eq. (S4), which will include phase information, but is beyond the scope of this paper.

## S2. Thermomechanical and Optical Shot Noise

The Langevin force is defined as $F_L = \sqrt{4k_B T c}\,\Gamma(t)$, where $k_B$ is Boltzmann's constant, $T$ is temperature, and $\Gamma(t)$ is a Gaussian white noise process with a standard deviation of 1[1]. Returning to eq. (S2), ignoring $a_e$, and taking the power spectral density of $x_c$ results in

$$S_{xx}(\omega) = |G(i\omega)|^2 \frac{4k_B T \omega_0}{mQ} \quad \text{(S5)}$$

The thermomechanical noise in terms of displacement is then defined as $x_{th} = S_{xx}(\omega)^{1/2}$, or

$$x_{th}(\omega) = |G(i\omega)| \sqrt{\frac{4k_B T \omega_o}{mQ}} \quad \text{(S6)}$$

Recalling the conversion from displacement to acceleration, eq. (S4), the equivalent acceleration due to thermomechanical noise is then

$$a_{th} = \sqrt{\frac{4k_B T \omega_o}{mQ}} \quad (S7)$$

Interestingly, $a_{th}$ is only a function of the resonator parameters ($\omega_0$, $m$, and $Q$) and temperature, and not a function of frequency, meaning that the thermomechanical noise floor in terms of acceleration is flat.

In addition to thermomechanical noise, optical shot noise is the other fundamentally limiting noise source. The power spectral density of the optical shot noise is $S_{PP} = 2h\nu P_a/\eta$, where $h$ is Planck's constant, $\nu$ is the optical frequency of the laser, $P_a$ is the average power reaching the photodetector, and $\eta$ is the quantum efficiency of the photodetector. This can be converted to shot noise in terms of displacement using

$$x_s = g_{x/V} g_{V/i} R S_{PP}^{1/2} = g_{x/V} g_{V/i} R \sqrt{2h\nu P_a/\eta} \quad (S8)$$

The gain $g_{x/V}$ converts photodetector voltage to displacement and is discussed in Section 4, while $g_{V/i}$ and $R$ are the transimpedance gain and responsivity of the photodetector. Recalling eq. (S4), the shot noise in terms of acceleration is

$$a_s = g_{x/V} g_{V/i} R \sqrt{2h\nu P_a/\eta} \left|G(i\omega)\right|^{-1} \quad (S9)$$

Since the thermomechanical noise and shot noise are uncorrelated, they can be summed in quadrature to get the total noise equivalent displacement, $x_{NE}$, and acceleration, $a_{NE}$.

Unlike the thermomechanical displacement noise, $x_{th}$, the optical shot noise does not represent real resonator motion but rather, it is detection noise that is analytically referred to either displacement or acceleration. As a result, the best-case scenario for a resonator with fixed parameters ($\omega_0$, $Q$, $m$, $T$) is for the optical shot noise to be lower than the thermomechanical noise. In this situation, the optical readout will measure the motion of the resonator with minimal contribution from shot noise. This is shown in Fig. S2, where the calculated noise floor is presented for a resonator with parameters similar to those described in the experiments in the letter. Three different levels of shot noise are shown, where two are above the thermomechanical noise (dark blue, light blue) and one is below (red). When the shot noise is below the thermomechanical noise, the resonance shape is observed over the entire frequency range, which improves the parameter fit when using the noise to determine the resonator parameters.

After converting the displacement to acceleration, as shown in Fig. S2b, the importance of reducing the shot noise is readily apparent. The noise equivalent acceleration is nearly flat over the frequency range when the shot noise is below the thermomechanical noise. Achieving a flat noise floor in acceleration is critical for a broadband accelerometer because it enables the measurement of signals with widely varying frequencies at the same precision level. For example, if the acceleration is a square wave, all of the harmonics within the bandwidth of the sensor will be measured with the same precision when the noise floor is flat, which means that the signal can be accurately reconstructed from the data. If the noise floor is frequency dependent, this reconstruction would be extremely difficult since the signal-to-noise ratio will vary across the frequency range.

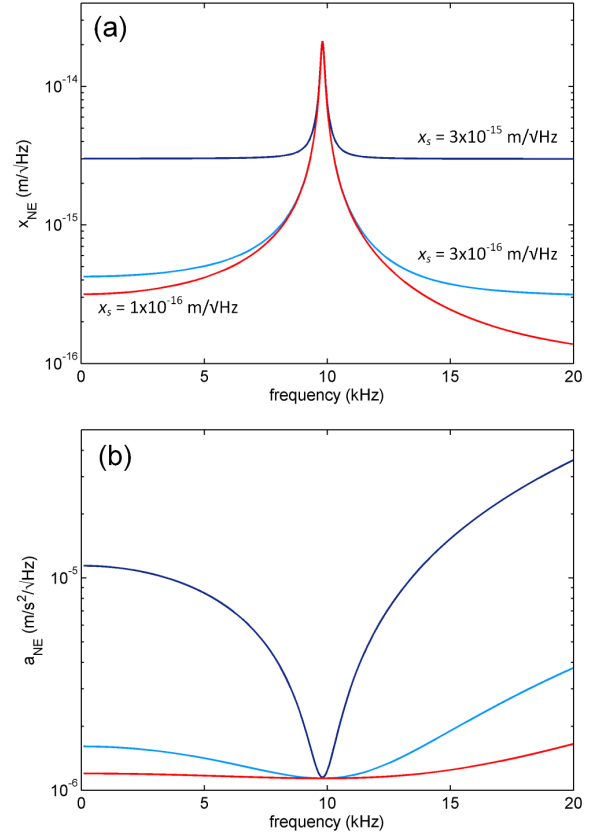

**Fig. S2 Noise equivalent displacement and acceleration for varying optical shot noise level.** (a) Noise equivalent displacement combining thermomechanical noise and optical shot noise at three different shot noise levels. $\omega_0 = 2\pi$ (9.8 kHz), $Q = 70$, $m = 11$ mg, $T = 293$ K. (b) Noise equivalent acceleration based on the displacement noise in (a).

It is also useful to look at the effect of $Q$ on the noise floor, as shown in Fig. S3. The shot noise is constant for these calculations while three values of $Q$ are used. The values of $Q$ range from setting the thermomechanical noise above the shot noise at all frequencies ($Q = 70$, dark blue line) to having the thermomechanical noise at low frequency to be equivalent to the shot noise ($Q = 7000$, red line). Lower $Q$ results in higher thermomechanical noise away from resonance, so that the true mechanical response of the resonator is more easily observed in the noise spectrum. After converting the displacement noise to a noise equivalent acceleration, as shown in Fig. S3b, the effect of increasing $Q$ becomes clearer. The noise equivalent acceleration for $Q = 7000$ is 10 times lower near resonance than when $Q = 70$, as expected from eq. (S7). However, the reduction in the noise floor is far less prominent away from resonance. At low frequency, the reduction is as low as a factor of 3 and it is much lower at frequencies above resonance. These modest improvements in the noise floor require an increase in Q by a factor of 100. Note that the trends seen in Fig. S3b are similar to the experimental results in Fig. 3 of the letter, thereby validating the models presented in this section.

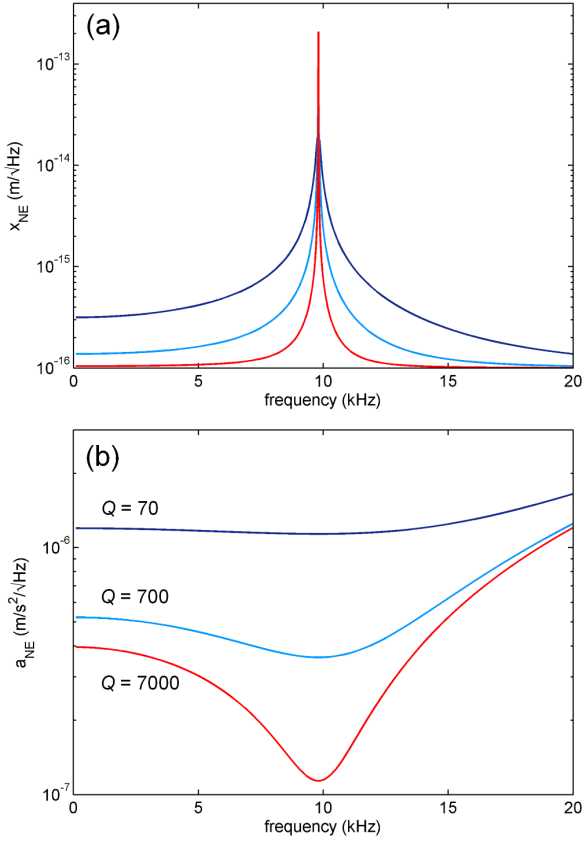

**Fig. S3 Noise equivalent displacement and acceleration for varying Q.** (a) Noise equivalent displacement combining thermomechanical noise and optical shot noise at three different Q values, 70, 700, and 7000. $\omega_0 = 2\pi$ (9.8 kHz), $m = 11$ mg, $T = 293$ K, $x_s = 1 \times 10^{-16}$ m/√Hz. (b) Noise equivalent acceleration based on the displacement noise in (a).

In summary, there are two important conclusions from this analysis. First, the optical shot noise should be lower than the thermomechanical noise. This will ensure that the acceleration noise floor is flat over the frequency range of interest, making it possible to perform broadband measurements with the same precision over that frequency range. It will also provide better estimates of $\omega_0$ and $Q$ when fitting displacement noise spectra to the harmonic oscillator model. Second, increasing $Q$ will only be useful for broadband measurements if the shot noise is lower than the thermomechanical noise over the entire frequency range of interest. Otherwise the reduction in the acceleration noise floor will be modest for large increases in $Q$ and will be offset by an increase in ringdown time and a reduction in the linear dynamic range, as discussed in the letter. Noting that the equation of motion in eq. (S2) also describes optomechanical sensors for force and pressure measurements, the above conclusions apply equally to these other domains.

## S3. Design of the Mechanical Resonator

The mechanical resonator has a large square single-crystal silicon proof mass (thickness: 525 μm, width: 3.02 mm (Device A) or 4.02 mm (Device B)) that is supported by an array of 1.5 μm thick silicon nitride beams, as shown in Fig. 1 of the letter. These beams are located around the entire perimeter of the proof mass and on both sides of the chip, where the beam length is selected to achieve the desired stiffness. This design increases the resonance frequencies for rotational modes of the proof mass (i.e., rocking modes) so that there is a large separation in frequency between the first translational mode (i.e., piston mode) and the other vibrational modes.

Structural finite element analysis (FEA) was performed for the two designs (Devices A and B) to assess the effectiveness of mode separation due to the flexural constraints. Figure S4 shows representative mode shapes for the first piston mode and first rocking mode. The piston mode is the mode of interest for detecting accelerations perpendicular to the chip surface. This mode exhibits pure translation of the proof mass along the optical axis, such that proof mass displacement causes a length change of the optical cavity. It was found that the resonance frequency of the first rocking mode is higher than the piston mode by a factor of 11.6 for Device A and 7.8 for Device B. This mode separation is sufficient to ensure that the rocking mode does not appear within the measurement bandwidth used for Fig. 3 in the letter. The closest mechanical mode detected in experiments is above 60 kHz, or a factor of 6 higher than the piston mode, as shown in Fig. S5b.

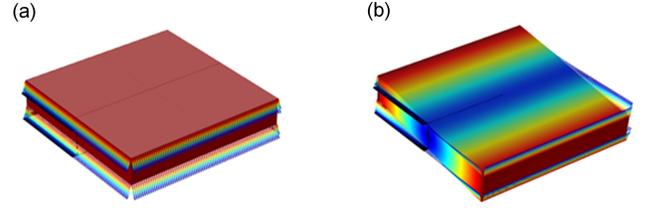

**Fig. S4 Mode shapes for the mechanical resonator.** (a) First piston mode, and (b) first rocking mode. Red indicates maximum displacement and dark blue represents no displacement.

## S4. Converting from Photodetector Voltage to Displacement

Displacement of the proof mass results in a change in cavity length, which is measured by the cavity readout. With the probing laser locked to the side of a TEM$_{00}$ optical resonance, the cavity length change, $\Delta L$, is transduced by measuring the change in the center wavelength of the optical resonance, $\Delta \lambda$, using:

$$\Delta L = \frac{L}{\lambda}\Delta\lambda \qquad (S10)$$

where $L$ is the nominal cavity length and $\lambda$ is the nominal laser wavelength at the lock point. The change in center wavelength, $\Delta\lambda$, is related to the reflected laser intensity from the cavity that is measured with a photodetector, resulting in a voltage change, $\Delta V$. The relationship between voltage and wavelength is defined by the slope of the optical resonance at the locking point, $dV/d\lambda$, as shown in the inset of Fig. S5a. The laser was locked to the point of greatest slope for the highest transduction sensitivity. In this way, the displacement of the proof mass is found using:

$$\Delta L = \frac{L}{\lambda}\Delta V \Big/ \left(\frac{dV}{d\lambda}\right) = g_{x/V}\Delta V \qquad (S11)$$

The parameters ($L$, $\lambda$, $dV/d\lambda$) are directly found from a spectral measurement of the cavity over a full free spectral range (FSR) and the voltage change, $\Delta V$, is measured with an electronic spectrum analyzer (ESA).

## S5. Readout using the External Cavity Diode Laser

Two different lasers were used for cavity readout: a continuously tunable external cavity diode laser (ECDL) and a tunable fiber laser (FL) that is phase modulated with an electro-optic modulator (EOM). The ECDL has a wide wavelength tuning range and precise piezo-based wavelength control, allowing for cavity characterization and FSR measurements, as shown in Fig. 2 of the letter. In comparison, the FL has a slow tuning rate and a much narrower tuning range. Furthermore, the internal feedback locking module of the ECDL enables direct and convenient cavity displacement readout. However, the ECDL has more internal frequency noise than the FL, which appears as noise equivalent displacement. Therefore, the FL was used for the displacement noise floor measurements in Fig. 3 of the letter since it has a cleaner frequency spectrum. Details on the readout method using the FL are described in the letter. Here, we provide additional information on the readout with the ECDL.

As shown in Fig. S5a, the main differences between using the ECDL and FL are the wavelength tuning method and the feedback servo loop. Wavelength tuning with feedback is achieved in the ECDL with a piezoelectric actuator in the external cavity. Therefore, unlike the FL, an EOM is not needed for locking. Regarding the implementation of the servo, the ECDL has an internal digital proportional-integral-derivative (PID) feedback controller while the FL servo uses an external analog PID controller.

A comparison of the displacement noise spectra from the accelerometer is shown in Fig. S5b for both readout lasers. No mechanical resonances other than the fundamental near 10 kHz are observed in the accelerometer up to 60 kHz. In general, the responses from the two lasers are very similar. However, the ECDL exhibits several resonances near 1.3 kHz that were determined to be mechanical resonances within the external cavity of the laser. The measurements in Fig. 4 were performed with the ECDL since the resulting displacements are well above the noise floor and the ECDL provides wider tuning range and simpler operation.

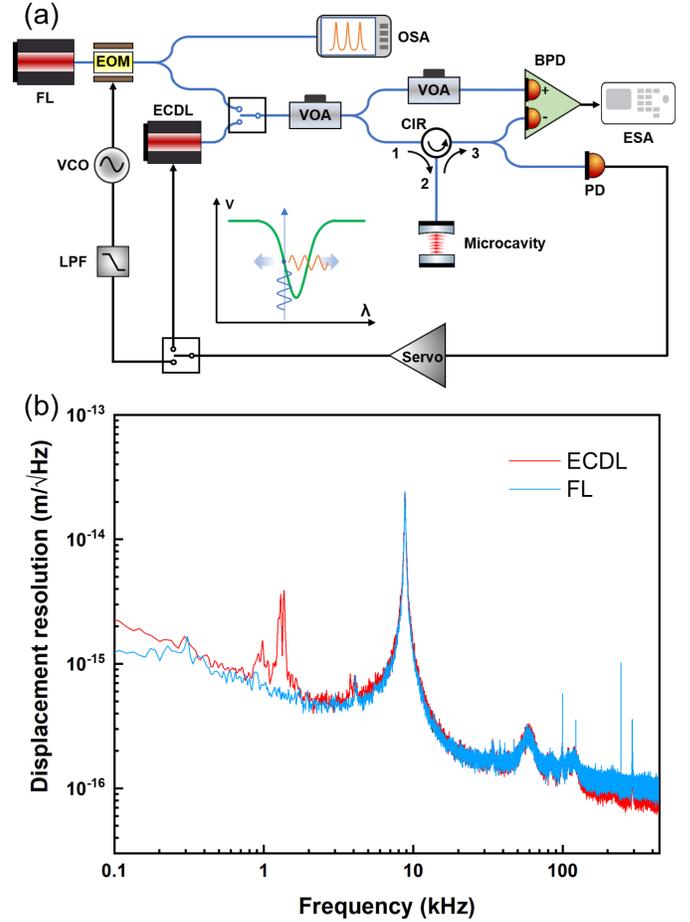

**Fig. S5 Cavity readout with the external cavity diode laser.** (a) Schematic of the cavity readout for the accelerometer using both the external cavity diode laser (ECDL) and fiber laser (FL). EOM: electro-optical modulator, SW: switch, OSA: optical spectrum analyzer; CIR: circulator, BPD: balanced photodetector, PD: photodetector, VOA: variable optical attenuator, ESA: electronic spectrum analyzer, LPF: low-pass filter, VCO: voltage-controlled oscillator. (b) Displacement noise spectra for the accelerometer when using the ECDL and FL.

## S6. Linearity of the Shaker Table

The comparison between the accelerometer and laser interferometer shown in Fig. 4 of the letter required that the excitation amplitude of the shaker be different when using the two measurement methods. This was due to the higher sensitivity of the accelerometer relative to the interferometer by a factor of approximately 600. As a result, higher excitation amplitudes were required for detection with the interferometer. These high excitation amplitudes could not be used while reading out the microcavity in the accelerometer because the side lock could not be maintained. The end result was that measurements with the interferometer were performed with excitation amplitudes that were as much as 50 times greater than with the accelerometer readout.

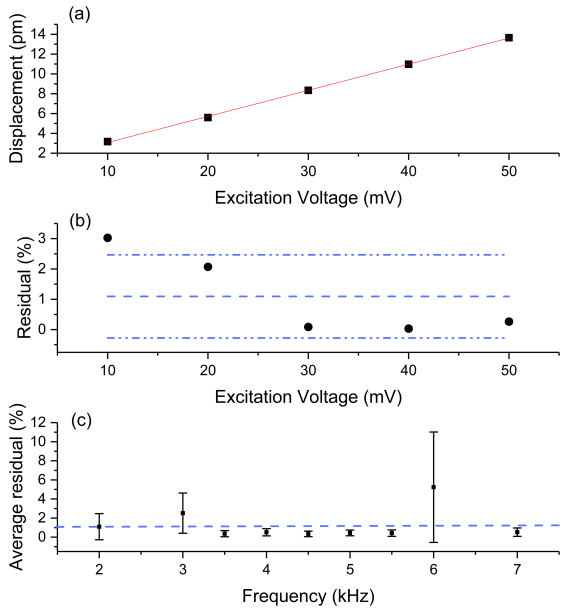

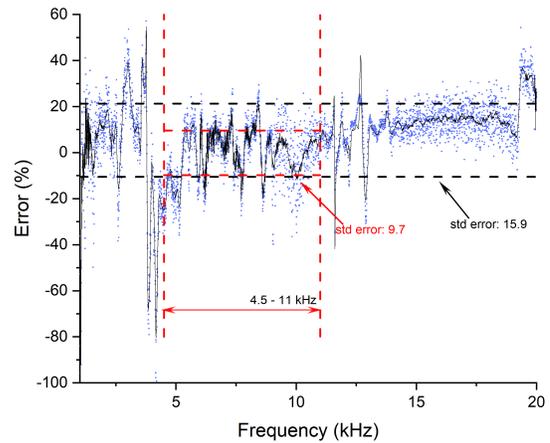

**Fig. S7. Comparison of the accelerometer and interferometer results on the shaker table.** Blue dots: deviation of the accelerometer results from the interferometer results. Black line: Same data set as blue dots but filtered using a moving average.

**Fig. S6. Linearity of the shaker table.** (a) Shaker table displacement as a function of excitation voltage at a drive frequency of 2 kHz. (b) Residuals from a linear fit to the data in (a). The residuals are an absolute value of the difference between the data and fit, expressed as a percentage of the fit value. Blue lines represent the mean (dash) and standard deviation (dash-dot) over the range of excitation voltages. (c) Mean and standard deviation residuals of the linear fit as a function of drive frequency. Blue line represents the mean over all frequencies.

This approach to the comparison is acceptable as long as the piezoelectric shaker table has a linear response for increasing excitation voltage. The linearity of the shaker table was characterized over a range of excitation voltages and frequencies, as shown in Fig. S6. The displacement of the shaker table for increasing excitation voltage at a single frequency (2 kHz) was found to be highly linear (Fig. S6a). The residuals for a linear fit to the data in Fig. S6a show a deviation from linearity of no more than 3 % and this deviation is much lower at higher excitation voltages due to the improved signal-to-noise ratio (Fig. S6b). Additional linearity measurements were performed between 2 and 7 kHz and the mean and standard deviation of the linear fit residuals were calculated (Fig. S6c). The shaker is linear within 3 % across the entire frequency range with the exception of an outlier at 6 kHz and the mean residual is 1.1 %. This level of linearity is more than adequate for the comparison between the accelerometer and interferometer, which is discussed further in the next section.

## S7. Accelerometer and Interferometer Comparison

The data in Fig. 4c of the letter was analyzed to compare the results from the accelerometer and interferometer when operating on the shaker table. The deviation of the accelerometer from the interferometer was calculated as a percentage, as indicated by the blue dots in Fig. S7. A moving average filter was applied to the data from the interferometer because noise in the data was found to be a major contributor to the deviation between the two measurements. This resulted in the black line in Fig. S7, showing a significant improvement in the comparison. The deviation for the filtered data is 5.4 % +/- 15.9 % (average +/- standard deviation) over the entire drive frequency range (1 kHz to 20 kHz). When looking at a narrower frequency range from 4.5 kHz to 11 kHz, the deviation is -0.1 % +/- 9.7 %. This deviation between accelerometer and interferometer is due to a number of factors but appears to be dominated by: 1) coupling between the shaker table and optomechanics in the interferometer, 2) dynamics of the stainless-steel package, and 3) the mounting interface. Each of these will be explored in future work.

**References**
S1. T.W. Gabrielson, Mechanical-thermal noise in micromachined acoustic and vibration sensors, *IEEE Trans. Electron Devices* **40**, 903–909, (1993).